\documentstyle[sprocl,epsf]{article}
\input{psfig}
\def\Journal#1#2#3#4{{#1} {\bf #2}, #3 (#4)}

\def\NPB{{\em Nucl. Phys.} B}
\def\PLB{{\em Phys. Lett.}  B}
\def\PRL{\em Phys. Rev. Lett.}
\def\PRD{{\em Phys. Rev.} D}
\def\ZPC{{\em Z. Phys.} C}

\def\be{\begin{equation}}
\def\ee{\end{equation}}
\def\bea{\begin{eqnarray}}
\def\eea{\end{eqnarray}}

\begin{document}
\title{Hadronic $B$ Decay}
\author{T. E. Browder}
\address{Physics Department, University of Hawaii at Manoa,\\ 
Honolulu, HI, 96822, USA}
%
%
\maketitle\abstracts{We review recent experimental
results from CLEO and LEP
experiments on hadronic decays of hadrons containing $b$
quarks\cite{bhp}. We discuss charm counting and the
semileptonic branching fraction in $B$ decays and
the color suppressed amplitude in $B$ decay.}

\section{Charm counting and the semileptonic branching fraction}

A complete picture of inclusive
$B$ decay is beginning to emerge from recent measurements 
by CLEO~II and the LEP experiments.\cite{bhp}
These measurements can be used to address
the question of whether the hadronic decay of the $B$ meson is compatible with
its semileptonic branching fraction.

Three facts emerge from the experimental examination of inclusive $B$
decay at the $\Upsilon (4S)$:
\begin{equation}  n_c = 1.10 \pm 0.05 \end{equation} 
where $n_c$ is the number of charm quarks
produced per $B$ decay from recent CLEO II
results\cite{cleo_bd}
 and using  
${\cal B}(D^0\to K^-\pi^+)=(3.91 \pm 0.08\pm 0.17\%)$.\cite{cleo_oldkpi}
\begin{equation}
{\cal B}(B\to X\ell\nu)=10.23\pm 0.39 \%.
\end{equation}
This value is the 
average of the CLEO and ARGUS model independent measurements using
dileptons.\cite{browhon} 
We note that the value used by the LEP Electroweak Working Group
for ${\cal B}(b\to X \ell\nu)=11.16\pm 0.20\%$ is only marginally
consistent with the $\Upsilon(4S)$ results.

The third quantity, ${\cal B}(b\to c\bar{c} s)$,
is calculated from the inclusive $B\to D_s$, 
$B\to (c\bar{c}) X$, and $B\to \Xi_c$ branching fractions, and is
\begin{equation} {\cal B}(b\to c \bar{c} s)= 14.0 \pm 2.8 \% .
\end{equation}
 The above value is determined assuming no contribution from $B\to D$ decays, 
an assumption which can be checked using data and is discussed in further
detail below.

\begin{figure}[htb]
\centerline{\epsfysize 2.7 truein\epsfbox{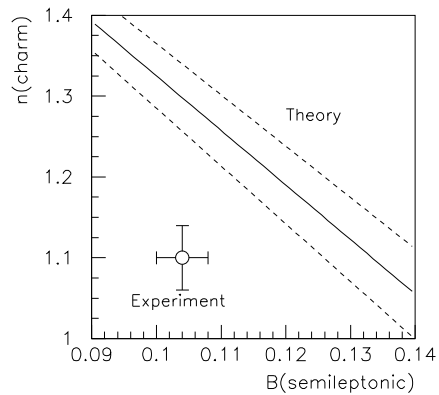}}
\caption{The value of $n_c$ versus the semileptonic $B$ branching
fraction using experimental values from $\Upsilon(4S)$
data. The measurements from LEP experiments are discussed in the text.}
\label{ncbsl}
\end{figure}


In the parton model, it is 
difficult to accomodate a low semileptonic branching fraction unless the 
hadronic width of the $B$ meson is increased.\cite{bsl}
The explanations for the semileptonic branching fraction
which have been proposed can be distinguished by expressing
the hadronic width of the $B$ meson in terms of three components:
$$\Gamma_{hadronic}(b)  = \Gamma (b\to c \bar{c} s) 
+ \Gamma  (b\to c \bar{u} d) +\Gamma (b\to s~g).$$ If the semileptonic
branching fraction is to be reduced to the observed level, then one of the
components must be enhanced.

A large number of explanations for the low semileptonic branching fraction
and charm yield have been proposed in the last few years.
These explanations can be logically classified as follows:
\begin{enumerate}
\item An enhancement of $b\to c \bar{c} s$ due to large QCD corrections
or the breakdown of local duality. A variety of possible 
experimental signatures
have been suggested.\cite{bccs1}$^{,}$\cite{bccs2}$^{,}$
\cite{bccs3}$^{,}$\cite{bccs4}$^{,}$\cite{bccs5}$^{,}$\cite{bccs6}

\item An enhancement of $b\to c \bar{u} d$ due to non-perturbative effects.
\cite{bcud1}$^{,}$\cite{bcud2}$^{,}$\cite{bcud3}$^{,}$\cite{bcud4}

\item An enhancement of $b\to s~g$ or $b\to d~g$ 
from New Physics.\cite{new1}$^{,}$\cite{new2}$^{,}$\cite{new3}

\item The cocktail solution: For example, 
if both the $b\to c \bar{c} s$ and the 
$b\to c\bar{u} d$ mechanisms are increased,
this could suffice to explain the inclusive observations.

\item There might also be a 
 systematic experimental problem in the determination of either $n_c$,
${\cal B}(b\to c \bar{c} s)$, or $ {\cal B}(B\to X \ell\nu)$.\cite{isisys}
\end{enumerate}


Inclusive charm particle-lepton correlations can be used to probe
the $B$ decay mechanism and give further insight into this problem. 
The correlation of the lepton charge and the charm particle flavor
distinguishes between different production mechanisms.
High momentum leptons, $p_{\ell}>1.4$ GeV, are used
to tag the flavor of the $B$. The angular correlation between the meson
and the lepton is then employed to select events in which the tagging lepton
and meson are from different $B$s. When the lepton and meson
originate from the same $B$ meson they tend to be back to back, whereas when
the meson and leptons come
from different $B$ mesons they are uncorrelated. 
After this separation is performed, wrong sign 
charge correlations from $B-\bar{B}$ 
mixing must be subtracted. Since the mixing rate is well measured, 
this correction is straightforward and has little uncertainty.

This technique has been applied previously to several types of 
correlations of charmed hadrons and leptons.
For example, the sign of $\Lambda_c$-lepton
correlations distinguishes between the 
$b\to c \bar{u} d$ and the $b\to c \bar{c} s$ mechanisms.

It was found that the $b\to c\bar{c} s$ mechanism comprises
$ 19\pm 13\pm 4 \%$ of $B\to \Lambda_c$ decays\cite{lamlep}. 
This observation effectively ruled out one proposed source of
additional $b\to c\bar{c} s$ decays.\cite{bccs1}
Similiarly, examination of the sign of 
$D_s$-lepton correlations shows that
most $D_s$ mesons originate from $b\to c \bar{c} s$ rather than
from $b\to c \bar{u} d$ with $s \bar{s}$ quark popping at the lower vertex.
In this case, it was found that $ 17.2 \pm 7.9 \pm 2.6\%$ of $D_s$ mesons
originate from the latter mechanism\cite{dslep}.
The same experimental 
technique has now been applied to $D$-lepton correlations.

The conventional $b\to c\bar{u} d$ mechanism which was {\it 
previously assumed} to be
responsible for all $D$ production in $B$ decay will give $D\ell^-$
correlations. 
If a significant fraction of $D$ mesons
arise from $b\to c\bar{c} s$ with light quark popping at the
upper vertex as proposed by Buchalla, Dunietz,
and Yamamoto significant wrong sign $D\ell^+$ correlations will be
observed.\cite{bccs2}

Final results of this study have been presented by CLEO~II which finds,
$\Gamma(B\to D~X)/\Gamma(B\to \bar{D} X) = 0.100\pm 0.026\pm 0.016$.\cite{kwon}
This implies a new contribution to the $b\to c \bar{c} s$ width
$${\cal B}(B\to D X) = 7.9\pm 2.2\%$$ ALEPH finds evidence for semi-inclusive
$B\to D^0\bar{D^0} X + D^0 D^{\mp} X$ decays with a somewhat larger branching 
fraction of $12.8\pm 2.7\pm 2.6 \%$.\cite{alephdd} 
DELPHI reports the observation of
$B\to D^{*+} D^{*-} X$ decays with a branching fraction 
of $1.0\pm 0.2\pm 0.3\%$.\cite{delphidd} 
Additional and quite compelling evidence that these
signals are due to $B\to D^{(*)}\bar{D}^{(*)} K^{(*)}$ decays
has been presented by CLEO\cite{cleoddk}, which has observed
fully reconstructed signals in exclusive modes:
$${\cal B}(\bar{B}^0\to D^{*+} \bar{D}^0 K^-) = 
0.45^{+0.25}_{-0.19}\pm 0.08\%$$
$${\cal B}(B^-\to D^{*0} \bar{D}^0 K^-) = 
0.54^{+0.33}_{-0.24}\pm 0.12\%$$
$${\cal B}(\bar{B}^0\to D^{*+} \bar{D}^{*0} K^-) = 
1.30^{+0.61}_{-0.47}\pm 0.27\%$$
$${\cal B}(B^-\to D^{*0} \bar{D}^{*0} K^-) = 
1.45^{+0.78}_{-0.58}\pm 0.36\%$$

The rates observed by ALEPH and DELPHI are 
consistent with the rate of wrong sign
$D$-lepton correlation reported by CLEO.\cite{kwon}
It is possible that these channels are actually resonant modes
of the form $B\to D {D}_s^{**} $
decays, where the p-wave $D_s^{**}$ or radially excited $D_s^{'}$ state 
decays to $\bar{D}^{(*)} \bar{K}$.\cite{blokddk} A direct
search by CLEO has ruled
out the possibility of narrow $B\to D_{s_1} X$ decay:
${\cal B}(B\to D_{s1}^+ X)<0.95\%$ at the 90\% confidence level.\cite{ds1}

There are other implications of these observations.
A $B$ decay mechanism with a ${\cal O}(10\%)$ branching 
fraction has been found
which was not previously included in the CLEO or LEP
Monte Carlo simulations of $B$ decay. This may have consequences
for other analyses of particle-lepton correlations. For example,
CLEO has re-examined the model independent dilepton measurement
of ${\cal B}(B\to X \ell\nu)$. Due to the lepton threshold of 0.6 GeV
and the soft spectrum of leptons, the CLEO measurement
is fortuitously unchanged. It is also important to check the size
of this effect in LEP measurements of the $B$ semileptonic branching fraction
using dileptons.

We can now recalculate\cite{kwon} 
$${\cal B}(b\to c \bar{c} s) = 21.9\pm 3.7 \% $$
which would suggest a somewhat larger charm yield ($n_c \sim 1.22$).
This supports hypothesis (1), large QCD corrections in $b\to c \bar{c} s$ 
{\it BUT the charm yield $n_c$ as computed in the usual way
is  unchanged}. Moreover, the contribution of 
$B\to D \bar{D} K X$ decays was properly accounted for
in the computation of $n_c$. This suggests that the experimental
situation is still problematic.

One possibility that must be addressed is whether 
there could be an error in 
the normalization ${\cal B}(D^0\to K^-\pi^+)$.\cite{isisys}
This branching fraction calibrates the inclusive measurements of
$B\to D^0$, $B\to D^+$, and $B\to D_s$ rates as well as
$n_c$. Historically, a flaw in 
${\cal B}(D^0\to K^-\pi^+)$ has been the culprit in other consistency
problems with charm counting.
\begin{table}[htb]
\begin{center}
\caption{Recent Measurements of ${\cal B}(D^0\to K^-\pi^+)$}
\medskip
\label{dkpi}
\begin{tabular}{ll}
Experiment & Measurement ($\%$) \\ \hline
ALEPH  &   $ 3.897\pm 0.094 \pm 0.117  $ \\ 
CLEO II  &   $ 3.91\pm 0.08\pm 0.17 $ \\
ARGUS  &   $ 3.41\pm 0.12\pm 0.28 $ \\
\end{tabular}
\end{center}
\end{table}
The most precise measurements of ${\cal B}(D^0\to K^-\pi^+)$
are obtained by fitting the $p_T$ spectrum
of soft pions in charm jets. An examination of Table~\ref{dkpi} shows
that these measurements are statistically precise but systematics
dominated. The Particle Data Group world average is currently dominated by
the ALEPH measurement.\cite{alephdkpi}
 Using $B$ decay data it is also possible to make consistency checks on the
calibration branching fraction. For example, the ratio
$$ {{\cal B}(B\to D^*\ell\nu)_{\rm partial}} \over 
{{\cal B} (B\to D^*\ell\nu)_{\rm full}} $$ where the decay in the
numerator is observed without reconstructing the $D$ decay gives
a measurement of the calibration branching fraction with very different
systematic effects. In CLEO data, this method
gives 
${\cal B}(D^0\to K^-\pi^+)= 3.81\pm 0.15\pm 0.16\%$.\cite{cleo_dkpi_partial}
Another quantity which can be examined is
$$ {{\cal B}(B\to D X \ell\nu) }\over{{\cal B}(B\to X\ell\nu)} $$
which should be unity modulo small corrections for semileptonic $D_s$ and
baryon production as well as for processes
which do not produce charm. Applied
to CLEO data, this method gives
${\cal B}(D^0\to K^-\pi^+)= 3.69\pm 0.08\pm 0.17\%$.\cite{kwon}
Neither method is sufficiently precise yet to 
conclusively demonstrate that either the
$D$ branching fraction scale is correct or that it has a systematic flaw.

Another possibility is enhanced ${\cal B}(b\to c u \bar{d})$.
On the theoretical side, Bagan et al. find that at next to leading order, 
$$ r_{ud} ={ {{\cal B}(b\to c \bar{u} d)}\over
{{\cal B}(b\to c \ell\nu)} } =4.0\pm 0.4$$
The value of ${\cal B}(b\to c \bar{u} d)$
 can be checked using measurements 
of inclusive $B$ decay from the $\Upsilon(4S)$ experiments:
$$ {\cal B}(b\to c \bar{u} d)_{exp} = {\cal B}(B\to D X) + 
{\cal B}(B\to \Lambda_c X)$$
$$ - {\cal B}(B\to D D_s X) - 2 {\cal B}(B\to D \bar{D} K X)          
-2.25 {\cal B}(b\to c\ell\nu) $$
$$ = (0.871\pm 0.035) + (0.036\pm 0.020) $$
$$ - (0.10\pm 0.027) -
2\times (0.079\pm 0.022) - (0.236\pm 0.010) $$
$$ {\cal B}(b\to c \bar{u} d)_{exp} =  0.41\pm 0.07 $$
In the above calculation, a small correction ($0.004$) has been applied
to the $B\to \Lambda_c X$ branching fraction to account
for $b\to c \bar{c} s$ production in baryonic $B$ decay. 
The factor of $2.25$ accounts for phase space suppression in
$b\to c\tau\nu$ decay. 
The experimental result is consistent with the theoretical
expectation,
$$ {\cal B}(b\to c \bar{u} d)_{theory} = 0.42\pm 0.04 $$
However, the present experimental accuracy is not quite sufficient
to completely rule out $b\to c\bar{u} d$  as the cause of the discrepancy.

We note that ALEPH and OPAL have recently reported a value for 
$n_c$ in $Z\to b\bar{b}$ decay.\cite{alephnc},\cite{opalnc} 
ALEPH finds $n_c^Z = 1.230\pm 0.036 \pm 0.038 \pm 0.053$.
The rate of $D_s$ and $\Lambda_c$ production is significantly higher
than what is observed at the $\Upsilon(4S)$. It is not clear whether
the quantity being measured is the same as $n_c$ at the $\Upsilon(4S)$, which
would be the case if the spectator model holds and if the contribution
from the other $b$-hadrons, $B_s$ and $\Lambda_b$, could be neglected. 
OPAL reports a somewhat lower value of $n_c =1.10\pm 0.045\pm 0.060\pm 0.037$
after correcting for unseen charmonium states. 
OPAL assumes no contribution from $\Xi_c$ production while ALEPH includes
a very large contribution from this source. The
contribution of $B\to $baryon decays to charm counting 
as well as the $\Lambda_c$, $\Xi_c$ branching fraction scales are still poorly
measured and definitely merit further investigation.

\section{Exclusive Hadronic Decays}


Recent progress has been made on partial reconstruction of hadronic
$B$ decays.\cite{dstpi} 
For example, the decay chain 
$$B\to D^* \pi_f, D^*\to (D) \pi_s$$
can be measured without reconstructing the $D$ meson. In this reaction,
there are five particles ($B, D^*, D, \pi_s, \pi_f$) with five 4-momenta
give 20 unknowns. The 4-momenta of the 
$\pi_s,\pi_f$ are measured which gives 8 constraints.
The $B, D, D^*$ masses and beam energy are known  and gives 4 constraints. 
Then energy-momentum conservation in the $B\to D^* \pi_f$ and $D^*\to D\pi_s$
decay chains gives 8 additional constraints. Thus, one can 
perform a $20-8-8-4=0$ C fit.

Two variables are used to extract the signal:
$\cos \Theta_D^*$, the angle between the $p_{\pi_s}$ and $p_B$
in the $D^*$ rest frame, and
$\cos\theta_B^*$, the angle between the $p_{\pi_F}$ and $P_B$
in the $B$ rest frame.

This method gives the most precise measurements of two exclusive branching
fractions:
$${\cal B}(\bar{B}^0\to D^{*+}\pi^-)=(2.81\pm 0.11\pm 0.21\pm 0.05)
\times 10^{-3}$$
$${\cal B}(B^-\to D^{*0}\pi^-)=(4.81\pm 0.42\pm 0.40\pm 0.21)
\times 10^{-3}.$$
The second systematic error is from the $D^*$ branching fractions.
A similiar partial reconstruction
analysis has been applied to the $B^-\to D^{**}(2420)^0\pi^-$
and $B^-\to D^{**}(2460)^0\pi^-$ decay modes.\cite{ddstpi} 
The event yields from fitting these distributions are substantial:
 $281\pm 56$ $D^{**}(2420)$, $165\pm 61$ $D^{**}(2460)$, although
there are also large background subtractions. These correspond
to branching fractions,
$${\cal B}(B^-\to D_1(2420) \pi^-) = (1.17\pm 0.24\pm 0.16\pm 0.03) 
\times 10^{-3}$$
$${\cal B}(B^-\to D_2^* (2460) \pi^-) = (2.1\pm 0.8\pm 0.3\pm 0.05) 
\times 10^{-3}$$
The former mode was previously
observed using a similiar technique by ARGUS. The latter mode is
observed for the first time by CLEO. As noted by J. Gronberg and
H. Nelson, the partial reconstruction technique may also be useful for
observing a time dependent CP asymmetry in $\bar{B}^0\to D^{*+}\pi^-$.

\subsection{The sign of the color suppressed amplitude and lifetimes}

The sign
and magnitude of the color suppressed amplitude can be determined using
several classes of decay modes in charm and bottom mesons. The numerical
determination assumes factorization and uses form factors from various
phenemonological models.

For $D$ decay one
uses exclusive modes such as $D\to K\pi$, $D\to K\rho$ etc., 
and obtains $$ a_1 = 1.10\pm 0.03,~ a_2 = -0.50\pm 0.03  $$
The destructive interference observed in two body $D^+$
decays leads to the $D^+$-$D^0$ lifetime difference.\cite{bhp}

For $B$ decay, one can find the magnitude of $|a_1|$ from
the branching fractions for the decay modes
$\bar{B}^0\to D^{(*)+}\pi^-$, $\bar{B}^0\to D^{+(*)}\rho^-$.
This gives $|a_1|=1.06\pm 0.03\pm 0.06$.
One can also extract $|a_1|$
from measurements of branching fractions
$B\to D^{+,(0)} D_s^{(*)-}$\cite{rodriguez}. 
The magnitude $|a_2|$ can be determined from the branching
fractions for $B\to \psi K^{(*)}$. This yields $|a_2|=0.23\pm 0.01\pm 0.01$.

The value of $a_2/a_1$ can be found by comparing
$B^-$ decays where both the external and spectator diagrams
contribute to $\bar{B}^0$ decays where only the external spectator
decays contribute. For example,
the model of Neubert et al. predicts the following ratios:
\begin{equation}
R_1 = {{\cal B}(B^- \to D^0 \pi^-) \over {\cal B}(\bar{B^0}\to D^+ \pi^-)}
                = (1 + 1.23 a_2/a_1)^2  \label{colrate1}
\end{equation}
\begin{equation}
R_2 = {{\cal B}(B^- \to D^0 \rho^-)
\over {\cal B}(\bar{B^0} \to D^+ \rho^-)}
                = (1 + 0.66 a_2 /a_1)^2  \label{colrate2}
\end{equation}
\begin{equation}
R_3 = {{\cal B}(B^- \to D^{*0} \pi^-)
         \over {\cal B}(\bar{B^0} \to D^{*+} \pi^-)}
                     =(1 + 1.29 a_2/a_1)^2  \label{colrate3}
\end{equation}
\begin{equation}
R_4 = {{\cal B}(B^- \to D^{*0} \rho^-)
          \over{\cal B}(\bar{B^0} \to D^{*+} \rho^-)}
                     \approx (1 + 0.75 a_2/a_1)^2   \label{colrate4}
\end{equation}

\begin{figure}[htb]
\centerline{\epsfysize 3.9 truein\epsfbox{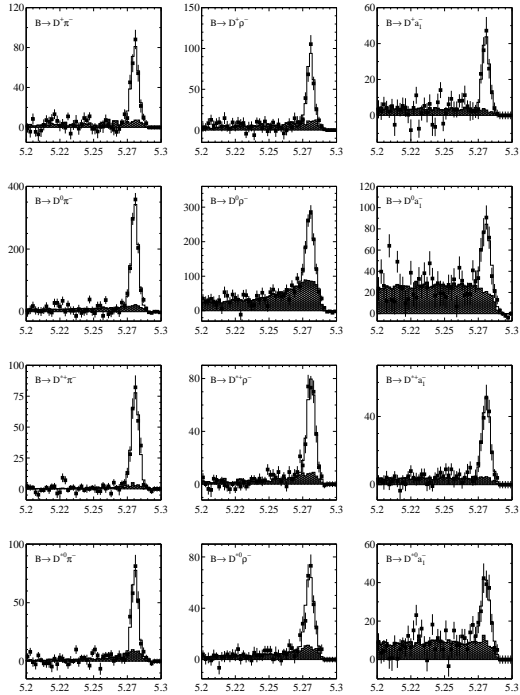}}
\caption{The beam constrained mass distributions of exclusive
hadronic decay modes used in the determination of $a_2/a_1$.
The mass plots have been continuum subtracted. The shaded
histogram is a high statistics
simulation of $B\bar{B}$ backgrounds.}
\label{bexclusive_multi}
\end{figure}

Improved measurements of these exclusive branching fractions
with better background subtraction 
and additional data have recently been 
presented by CLEO (see Fig.~\ref{bexclusive_multi}).\cite{rodriguez}
Using the latest branching fractions, 
$$ a_2/a_1 = 0.21 \pm 0.03 \pm 0.03^{+0.13}_{-0.12},$$ where the third
error is a conservative estimte of the uncertainty ($\sim 20\%$)
in the relative production of $B^+$ and
$B^0$ mesons at the $\Upsilon(4S)$. There are a number of 
additional theoretical uncertainties
which could significantly modify the magnitude of $a_2/a_1$ but
not its sign. For example,
the ratios of some heavy-to-heavy to heavy-to-light form factors is needed
(e.g. $B\to \pi/B\to D$). 
Comparing the value of $a_2/a_1$ determined using
form factors from the model of Neubert {et al.} with the value obtained
using form factors from the
model of Deandrea {\it et al.} shows that this uncertainty is small.
The effect
of including the $B\to V V$ mode for which the form factors have somewhat 
larger theoretical uncertainties is also small. 
It is important to remember that
the determination of $a_2/a_1$ also assumes the factorization hypothesis.
The large error on the relative production of $B^+$ and $B^0$ mesons
is the most significant experimental
uncertainty in the determination of $a_2/a_1$. 

The value of $a_2/a_1$ determined above
is consistent with the ratio $|a_2|$/$|a_1|$ where $|a_2|$ is computed
from $B\to \psi$ modes and $|a_1|$ is computed from $\bar{B}^0\to D^{(*)}\pi,
D^{(*)}\rho$ modes. Although the result is surprisingly different
from what is observed in hadronic charm decay (where the interference
is destructive) and from what is expected in the $1/N_c$ expansion,
Buras claims that the result can be accomodated
in NLO QCD calculations.\cite{buras}

If the constructive interference which is observed in these
$B^+$ decays is present in all $B^+$ decays, then we expect
a significant $B^+$-$B^0$ lifetime difference ($\tau_B^{+}< \tau_{B^0}$), 
of order $15-20\%$, in a direction opposite
to the $D^+-D^0$ lifetime difference. This
scenario is only marginally consistent 
with experimental measurements of lifetimes;
the world average computed by the LEP lifetime working group  
in August 1997 is $$\tau_{B^+}/\tau_{B^0}= 1.07 \pm 0.04 $$


It is possible 
that the hadronic $B^+$ decays that have been observed to date are 
atypical. The remaining higher multiplicity $B^+$ decays could
have destructive interference or no interference.\cite{neubert2bd} 
Or perhaps there is a mechanism which also enhances the $\bar{B}^0$
width to compensate for the increase in the $B^+$ width
and which maintains the $B^+/B^0$ lifetime ratio near unity.
Such a mechanism would be relevant to the charm counting and
semileptonic branching fraction problem.
In either case, there will be experimental consequences in the
pattern of hadronic $B$ branching fractions.\cite{lipkin}  
Experimentally one can compare other $B^-$ and $B^0$ decays including 
$D^{**}\pi^-$ and $D^{**}\rho^-$ as well
decays to $D^{(*)} a_1^-$, $a_1^-\to \rho^0\pi^-$ 
and $D^{(*)} b_1^-$, $b_1^-\to \omega\pi^-$ to check the first
possibility.

\section{Conclusion}

The charm counting and semileptonic $B$ branching
fraction problem persists. Three possible
solutions are still experimentially viable. 
These are (1) a systematic problem in the $D$
branching fraction, (2) an enhancement of $b\to c\bar{u} d$ or 
(3) an enhancement of $b\to s g$. Any proposed solution must also satisfy
the experimental constraints on ${\cal B}(b\to c \bar{c} s)$ and
${\cal B}(b\to c \bar{u} d)$. 

The sign of $a_2/a_1$ is found to be positive in the low multiplicity hadronic
$B$ decays that have been observed so far. This indicates
constructive interference in hadronic $B^+$ decay.
It will be interesting to see whether this pattern persists
as higher multiplicity $B$ decay modes are measured.

\end{document}